\documentclass[sigconf]{acmart}

\usepackage{multirow}
\usepackage{booktabs,siunitx}

\usepackage{amsmath,amsfonts,amsthm}
\usepackage{graphicx}
\usepackage{caption}
\usepackage{subcaption}

\AtBeginDocument{%
  \providecommand\BibTeX{{%
    \normalfont B\kern-0.5em{\scshape i\kern-0.25em b}\kern-0.8em\TeX}}}

\newcommand\red[1]{\textcolor{red}{#1}}
\newcommand\green[1]{\textcolor{green}{#1}}

\def\BibTeX{{\rm B\kern-.05em{\sc i\kern-.025em b}\kern-.08em
    T\kern-.1667em\lower.7ex\hbox{E}\kern-.125emX}}

\copyrightyear{2024}
\acmYear{2024}

\acmConference[KDD'24]{ACM}{Aug 25 -- 29, 2024}{Barcelona, Spain}
\acmPrice{15.00}
\acmISBN{978-1-4503-XXXX-X/18/06}

\begin{document}

\title{General Item Representation Learning for Cold-start Content Recommendations}

\author{Jooeun Kim}
\email{kje980714@snu.ac.kr}
\affiliation{
  \institution{Seoul National Univ.}
  \country{Seoul, Korea}
}

\author{Jinri Kim}
\email{ruth9811@snu.ac.kr}
\affiliation{
  \institution{Seoul National Univ.}
  \country{Seoul, Korea}
}

\author{Kwangeun Yeo}
\email{kwangeun.yeo@snu.ac.kr}
\affiliation{
  \institution{Seoul National Univ.}
  \country{Seoul, Korea}
}

\author{Eungi Kim}
\email{kuman5262@snu.ac.kr}
\affiliation{
  \institution{Seoul National Univ.}
  \country{Seoul, Korea}
}

\author{Kyoung-Woon On}
\email{kloud.ohn@kakaobrain.com}
\affiliation{
  \institution{Kakao Brain}
  \country{Sungnam, Korea}
}

\author{Jonghwan Mun}
\email{jason.mun@kakaobrain.com}
\affiliation{
  \institution{Kakao Brain}
  \country{Sungnam, Korea}
}

\author{Joonseok Lee}
\email{joonseok@snu.ac.kr}
\affiliation{
  \institution{Seoul National Univ.}
  \country{Seoul, Korea}
}

\settopmatter{authorsperrow=4}

\begin{abstract}
Cold-start item recommendation is a long-standing challenge in recommendation systems. A common remedy is to use a content-based approach, but rich information from raw contents in various forms has not been fully utilized. In this paper, we propose a domain/data-agnostic item representation learning framework for cold-start recommendations, naturally equipped with multimodal alignment among various features by adopting a Transformer-based architecture. Our proposed model is end-to-end trainable completely free from classification labels, not just costly to collect but suboptimal for recommendation-purpose representation learning. From extensive experiments on real-world movie and news recommendation benchmarks, we verify that our approach better preserves fine-grained user taste than state-of-the-art baselines, universally applicable to multiple domains at large scale.
\end{abstract}


\maketitle

\section{Introduction}
\label{sec:intro}

Recommendation systems are widely adopted for a variety of {real-world} applications, \emph{e.g.}, online retails, video sharing platforms, and more, as the scale of items that people may choose from has been rapidly growing. Collaborative filtering (CF)~\cite{Goldberg1992tapestry,Breese1998cf}, recognizing preference patterns observed in user-item interactions, has been successfully applied to personalized recommendation systems to provide potentially preferred items in a personalized manner.

Despite its success, CF approaches suffer from several challenges, one of which is the cold-start problem.
Since CF relies only on user and item interaction, it is not capable of generating personalized recommendations for a new user without any records. Likewise, a brand-new item with no user feedback cannot be recommended to the right customers who are most likely to prefer that item.

Cold-start is actually a common problem in modern recommendation systems. On YouTube, for example, 500 hours of contents are being uploaded every minute~\cite{Hwang2019flume}.
With a CF recommendation system, fresh contents can only be recommended to some random users until sufficient interaction data is collected.
Another example is Netflix, where new movies or TV series often compete for a limited main advertisement space. It is important for the supplier to select users who will most likely enjoy the new contents to maximize its revenue, where cold-start item recommendation plays a key role in selecting the right set of users for each fresh content without any user feedback.
Another domain that the cold-start is important is news articles. Unlike multimedia contents that their value lasts for a long time, news contents are useful only for a short period. In other words, it is more important to recommend news articles to the right people before we collect sufficient activities on them, and thus cold-start is the key in this domain.

To tackle the cold-start problem, side information about the users or items has been utilized.
Since content information becomes available at the time of release, it is possible to retrieve a set of neighboring items that are of similar content, and it may be recommended to users who like this kind of items.
Traditional approaches~\cite{zhao2022cvar,liu2019recommender,ning2012sparse,sun2019research} used demographic information of the users or meta-data of the items, \emph{e.g.}, genre or artist, to get prior knowledge of them.
With recent advances in deep learning, extracting semantics from the raw content, \emph{e.g.}, videos~\cite{tran2018r3d,carreira2017i3d,arnab2021vivit} or music~\cite{park2017representation,jiang2020transformer,huang2022mulan}, has become pervasive.
Some recent works attempt to learn more powerful item representations for recommendation purpose, directly from the raw contents~\cite{Lee2018cdml,Lee2020gcml,wei2021clcrec}.

Here, we pose the key question: is this rich content information being properly and sufficiently utilized for cold-start recommendations?
From two observations below, we believe it is still limited.

First, most existing methods are specifically designed for a particular dataset on a specific domain.
This is probably because the ``content information'', by nature, varies depending on the domain, service, or dataset.
For instance, movie contents densely provide visual frames, while music contents are mainly sound tracks. A visual-signal-based movie recommendation model will not work for music due to the absence of visual signals. The opposite, a sound-track-based music recommendation model may also not work well for movies, as audio in the movies is not as dense and informative as in music.
For this reason, content-based recommendation models have been developed independently for each domain and dataset, without having a general, common framework like CF.

Second, in spite of the advances in deep learning, raw multimedia contents are still not easily utilized due to their high cost of training and serving.
Specifically, it is a common practice to learn multimedia representations on large, human-labeled classification data, \emph{e.g.}, JFT~\cite{sun2017jft}, HowTo100M~\cite{miech2019howto100m} or Kinetics~\cite{carreira2017i3d}.
Setting aside the high cost of collecting a large volume of examples and labeling them, we claim that feature encoding learned from classification labels is sub-optimal for recommendations.
Classifying a sample to a predefined category intrinsically forces the model to learn only the common aspects within each class, ignoring subtle differences among individual examples in the same class.
We hypothesize that an encoding learned from such a classifier does not sufficiently preserve fine-grained details that are necessary and useful for a recommendation model to distinguish subtle preference of individual users on a variety of items.

In this paper, we seek a general item content representation learning framework which is domain and dataset-agnostic, and preferably, which does not rely on a human-labeled classification dataset.
In order to overcome the aforementioned problems, we propose to utilize the Transformer~\cite{Vaswani2017attention} architecture, which has been the basis of most state-of-the-art models in recent language~\cite{Devlin2018bert,ma2019tensorized}, image~\cite{dosovitskiy2020vit,liu2021swin}, video~\cite{arnab2021vivit,bertasius2021space} and audio~\cite{gong2021ast} understanding.
Transformer architecture is particularly appropriate for our use case, since it applies in most steps a common architecture to the input sequence regardless of its nature, once each input token is mapped to an embedding simply by a linear mapping.
In other words, its input-type-specific part is light, making it more appropriate for a general feature extractor.
This is in contrast to a CNN-based image model or an RNN-based text model, where the architecture-specific representations are kept until the very last few layers.
Thanks to its nature to rely less on data modality, it also provides a natural way of multimodal fusion~\cite{Lu2019vilbert,sun2019learning,akbari2021vatt}.
Our proposed framework is trained end-to-end solely on user activities, \emph{e.g.}, clicking or rating, without pre-training on classification labels on a predefined set.

To verify effectiveness and generalizability of the proposed approach, we conduct extensive cold-start recommendation experiments on multiple domains, where the multimodal content signals are particularly rich (movie) and where the cold-start recommendation is particularly important (news).
From experiments, we demonstrate that the content representations learned by our general framework perform significantly better recommendations than existing methods, preserving finer subtleties about items.
Note that we focus only on the cold-start recommendations, where the content signals are required and play the key role. Combining this with a CF-based model for warm-start cases will be an interesting extension, but this is beyond the scope of this paper.

Our main contributions are summarized as follows:
\begin{itemize}
  \item We propose a \emph{domain/dataset-agnostic} item content representation learning framework for cold-start recommendations, effectively fusing \emph{multimodal signals}.
  \item Our framework is \emph{end-to-end trainable}, without relying on human-labeled large-scale classification data to train modality-specific encoders. Trained solely on user activities, our item representations better preserve \emph{fine-grained taste} of users.
  \item From extensive experiments, we demonstrate that our proposed approach achieves state-of-the-art performance on cold-start recommendation on large-scale datasets from multiple domains.
\end{itemize}

\section{Related Work}
\label{sec:related}

\subsection{Cold-start Recommendations}
Collaborative Filtering (CF) has been successful in personalized recommendation systems with the existence of plentiful historical data~\cite{schafer2007collaborative,su2009survey,Lee2013llorma,he2017neural,rendle2009bpr,zhang2014collaborative}, but the cold-start problem is its long-standing challenge, where no historical interaction record of user or item exists. To tackle this problem,
MWUF~\cite{zhu2021learning} warms up cold items with meta-scaling and shifting networks.
DropoutNet~\cite{volkovs2017dropoutnet} randomly drops items or users to make the model better adapt to cold-start. 
Heater~\cite{zhu2020recommendation} tackles the problem with a randomized training mechanism and mixture-of-experts transformation.
Recently, meta-learning approaches~\cite{dong2020mamo,pan2019warm,lu2020meta,lee2019melu,yu2021personalized} are proposed to tackle cold-start recommendation.

\subsection{Content-based Recommendations}
Auxiliary information like content features has been integrated to CF models to alleviate the cold-start problem.
CB2CF~\cite{barkan2019cb2cf} connects the gap between item content and their CF representations.
CWH~\cite{barkan2021cold} balances the quality of warm and cold items by utilizing text-based item content.
CLCRec~\cite{wei2021clcrec} maximizes the mutual dependencies between item content and collaborative signals using contrastive learning.
CLCRec shares a common theme with our model in that it utilizes multimodal content features to tackle cold-start recommendation.
However, it trains embeddings on image classification labels and transfers them to the recommendation task, while our framework is completely free from human labels.
More recently, CVAR~\cite{zhao2022cvar} adopts conditional variational autoencoders to warm up cold item embeddings using content metadata. 
Recently, graph neural networks (GNN) become increasingly prevalent in recommender systems.
PMGT~\cite{wu2022graph}, for example, combines GNN with multimodal side information in item recommendation~\cite{liu2021pmgt}.
There are more examples, \emph{e.g.},
DUIF~\cite{geng2015learning}, MTPR~\cite{du2020learn},
CC-CC~\cite{shi2019adaptive},
MMGCN~\cite{wei2019mmgcn}, and Movie Genome~\cite{deldjoo2019movie}.
See a survey~\cite{deldjoo2020recommender} for more.

CDML~\cite{Lee2018cdml} is another model that proves usefulness of audio-visual features in cold-start scenario. GCML~\cite{Lee2020gcml}, learns video embeddings from a relational graph.
However, both models are not personalized in that they learn item-item co-watch similarity aggregated over all users, not at individual user level. On the other hand, our model explicitly uses individual user feedback to learn the item representations.

Although many content-based approaches tackle cold-start, they are restricted to a particular domain and features specific to the target dataset, often trained on classification labels. Our approach, on the other hand, is applicable to arbitrary domains and features, and is end-to-end trainable free from human-labeled data.

\subsection{News Recommendations}
News recommendation models particularly exploit content features to tackle the item cold-start problem~\cite{kompan2010content, wang2018dkn, wu2022mm}, since news articles are replaced with new ones in a short period of time.
For instance, NRMS~\cite{wu2019neural0} and NPA~\cite{wu2019npa} learn article representations from news titles. 
TANR~\cite{wu2019neural1}, LSTUR~\cite{an2019neural}, and NAML~\cite{wu2019neural2} utilize news topics or article bodies in addition to titles to enrich the news representations.
Most existing models exploit textual modality to represent news articles~\cite{wu2021empowering,wu2019neural0,wu2019neural1,wu2019neural2,an2019neural}, but recently, visual modality is also considered~\cite{wu2022mm}.
Along with this trend, our framework supports arbitrary number of content features in various forms, including visual and textual.
Recently, visual modality is also considered~\cite{wu2022mm}.
Along with this trend, our framework supports arbitrary content features, including visual and textual.

\subsection{Contrastive Learning}
Contrastive learning is a self-supervised task, learning to discriminate which pairs of data points are similar and different from the dataset, widely used in computer vision and NLP~\cite{chen2020simclr,grill2020bootstrap,he2020momentum,khosla2020supervised,hjelm2018learning,chi2020infoxlm}.
Recent works employ contrastive learning in recommender systems to optimize the user and item representations. 
For instance, Liu et al.~\cite{liu2021contrastive} proposes a graph contrastive learning to alleviate the sample bias.
CLRec~\cite{zhou2021contrastive} employs it to improve DCG in recommendation.
SLMRec~\cite{tao2022self} incorporates contrastive learning into multimedia recommendation with a graph neural network.
Our method also employs contrastive loss for rating prediction and multi-modal alignment, detailed in Sec.~\ref{sec:method}.

\section{Problem Formulation and Notations}
\label{sec:problem}

\begin{figure}
  \centering
     \centering
     \includegraphics[width=\linewidth]{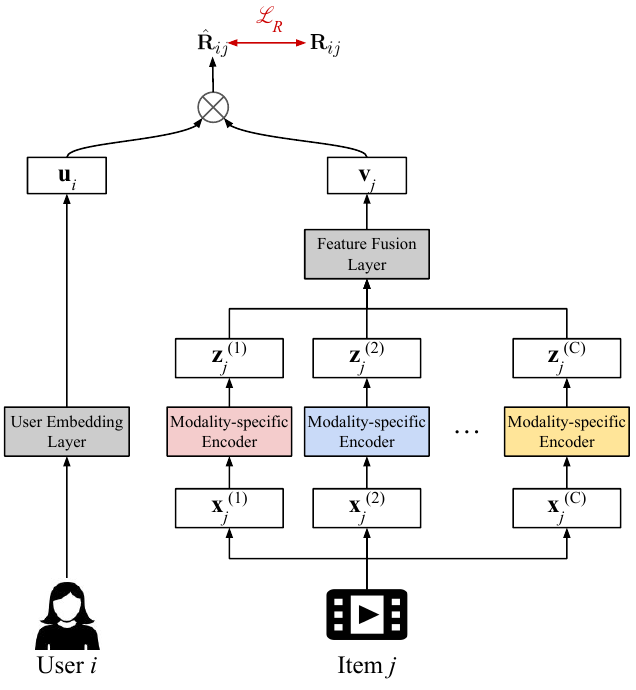}
     \caption{\textbf{Overall Architecture}. $C$ content features are extracted for each item using modality-specific encoders. (A few examples are illustrated in Fig.~\ref{fig:model_encoders}.) Then, the Feature Fusion Layer aggregates them into the final item representation $\mathbf{v}_j$, and the rating $\mathbf{R}_{ij}$ is predicted by taking dot product with the target user embedding $\mathbf{u}_i$, learned in the manner of collaborative filtering.}
     \label{fig:model_overall}
     \label{fig:overview}
\end{figure}
  
\begin{figure*}
    \centering
    \includegraphics[width=\linewidth]{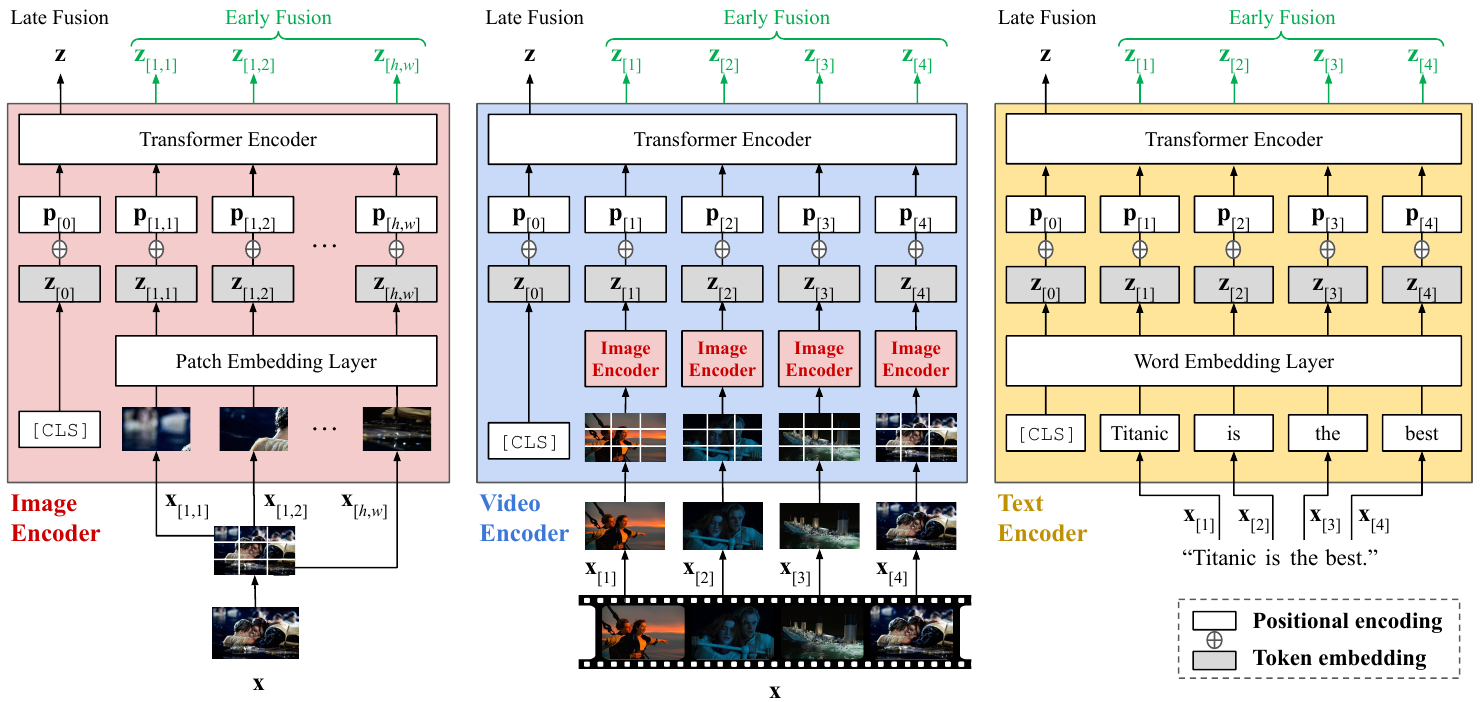}
    \caption{\textbf{Examples of Modality-specific Encoders}. From the left, we illustrate the image, video, and text encoders.}
    \label{fig:model_encoders}
\end{figure*}

In this paper, we presume implicit feedback from the users, so there are only two types of ratings: \emph{preferred} and \emph{unknown}.
Given a binary preference matrix $\mathbf{R} \in \{0, 1\}^{M \times N}$ with $M$ users and $N$ items, an element $\mathbf{R}_{ij} = 1$ indicates that the user $i$ prefers the item $j$, while $\mathbf{R}_{ij} = 0$ means unknown. The matrix $\mathbf{R}$ can be split into two parts: $\mathbf{R}_w$ with warm items and $\mathbf{R}_c$ with cold items, where all entries within $\mathbf{R}_c$ are zeros. The cold-start recommendation task is predicting preferable items within $\mathbf{R}_c$; in other words, retrieving a list of items that each user $i$ may prefer among the cold items.

Each item is provided with a set of $C$ content attributes. The content information for each attribute $c = 1, ..., C$ is denoted by $\mathbf{X}^{(c)} \in \mathbb{R}^{N \times D_c}$, where $D_c$ is the dimensionality of the content information for the attribute $c$.
Depending on its nature (modality), $D_c$ may be in a structured form. For an image (\emph{e.g.}, a raw frame), for instance, $D_c = H_c \times W_c \times 3$, where $H_c$ and $W_c$ are the height and width of the image. For a video, $D_c = T_c \times H_c \times W_c \times 3$, where $T_c$ is the maximum number of frames in the video. For a textual modality (\emph{e.g.}, synopsis), $D_c = \{1, ..., |V|\}^{T_c}$, where $T_c$ is the maximum length of the text for $c$ and $V$ is the vocabulary set.
The content information for $c$ of a particular item $j$ is denoted by $\mathbf{X}^{(c)}_j \in \mathbb{R}^{D_c}$.

We tackle cold-start items only, not cold-start users, since no public dataset provides meaningful user side information due to privacy, although cold-start users can be modeled in a similar way.

\section{Preliminary}
\label{sec:preliminary}

We briefly review Transformers~\cite{Vaswani2017attention}, on which our general item representation learning framework is built.
Transformer is a powerful model that achieves state-of-the-art performance on sequence-to-sequence tasks~\cite{Lu2019vilbert} like machine translation as well as general representation learning for images~\cite{dosovitskiy2020vit} and videos~\cite{arnab2021vivit}.
Taking as input a sequence of its sub-component (\emph{e.g.}, words for a sentence, smaller patches for an image, and frames for a video), it applies a self-attention mechanism in an encoder-decoder structure to learn context by tracking relationships among those sub-components.
We first describe the Transformer encoder in detail, followed by how it is utilized for two important modalities: text and visual. The decoder is not used in our framework.

\subsection{Transformer Encoder}
Recall that a side information $c$ for an item $j$ is denoted by $\mathbf{X}^{(c)}_j$. To be uncluttered, we omit $c$ and $j$ whenever clear.
$\mathbf{X}$ is split into a sequence of $T$ sub-components, denoted by $\{\mathbf{x}_{[1]}, ..., \mathbf{x}_{[T]}\}$, and how to split varies by modalities. Some modalities like a video or a sentence are sequential in nature. An image may be split into multiple smaller patches~\cite{dosovitskiy2020vit}.

Given this sequence $\{\mathbf{x}_{[1]}, ..., \mathbf{x}_{[T]}\}$ of $T$ tokens as input, they are first embedded into vectors, $\mathbf{Z} \equiv \{\mathbf{z}_{[1]}, ..., \mathbf{z}_{[T]}\}$, where $\mathbf{z}_{[t]} \in \mathbb{R}^d$ and $d$ is the token embedding size.
Then, $\mathbf{Z}$ is fed to a series of encoder blocks, where each block is composed of a self-attention layer and a feed-forward network, which enrich token representations with contextual information from other tokens in the sequence.

First, the token embeddings $\mathbf{Z} \in \mathbb{R}^{T \times d}$ are transformed to three special representations, namely, query ($\mathbf{Q} \in \mathbb{R}^{T \times d'}$), key ($\mathbf{K} \in \mathbb{R}^{T \times d'}$), and value ($\mathbf{V} \in \mathbb{R}^{T \times d'}$), by linear transformation, where $d'$ is not necessarily same as $d$.
Then, the self-attention is defined as
$\text{Attention}(\mathbf{Q}, \mathbf{K}, \mathbf{V}) = \text{softmax} \left( \mathbf{QK}^\top / \sqrt{d'} \right) \mathbf{V}$.
Intuitively, the attention of each token is represented as a weighted average of other token embeddings (using $\mathbf{V}$) in the same sequence, where the weight is proportional to the relevance (computed using $\mathbf{Q}$ and $\mathbf{K}$) between them.
The learnable parameters are linear mappers from token embeddings to $\mathbf{Q}$, $\mathbf{K}$, and $\mathbf{V}$.
Multiple heads are often used to allow each token to represent more than one semantics depending on the context.

After the multi-head self-attention, the embeddings are fed into a position-wise feed-forward network, allowing further transformation.
These steps are repeated by $L$ blocks.
The output of the last encoder block is the final embedding of each token.
Optionally, we may put an additional classification token (\texttt{[CLS]}) to learn the aggregated representation of the entire sequence.
Without having specific meaning, \texttt{[CLS]} aggregates tokens without being biased towards itself as other regular tokens do.
The Transformer is often trained by losses arisen from a downstream task like classification, performed based on this aggregated embedding from \texttt{[CLS]} token.

\subsection{Transformers for the Text Modality.}
Bidirectional Encoder Representations from Transformers (BERT)~\cite{Devlin2018bert} is a language model that learns representations from unlabeled text by self-supervised learning, based on the Transformer encoder.
The main training objectives are to predict masked tokens in sentences (Masked language modeling; MLM) and to predict whether two input sentences are consecutive (Next Sentence Prediction; NSP). With MLM, the randomly masked tokens are classified based on context (remaining tokens). For NSP, the embedding corresponding to the \texttt{[CLS]} token is fed to a classifier determining if the two input sentences are consecutive. For both, a classification loss (\emph{e.g.}, cross entropy) is used to train the model.
BERT is powerful in precisely learning semantics of words when trained on large-scale corpus, achieving state-of-the-art performance on various NLP tasks.

\subsection{Transformers for the Visual Modality.}
The Vision Transformer (ViT)~\cite{dosovitskiy2020vit} is a Transformer-based object recognition or image classification model. ViT employs a Transformer over fixed-size (\emph{e.g.}, 16 $\times$ 16) patches split from the input image. Each image patch is linearly transformed to a patch embedding, added with learnable positional encoding and fed into the Transformer encoder.
Optionally, multiple blocks of Transformers may be stacked.
At the end of the last block, a learnable classification \texttt{[CLS]} token is appended to aggregate the learned representation of the entire image.
It is fed into an MLP head performing the downstream task, \emph{e.g.}, image classification.
Video Vision Transformer (ViViT)~\cite{arnab2021vivit} and TimeSFormer~\cite{bertasius2021space} extend this idea to the sequence of frames, equipped with several options to reduce computational overhead.

\section{The Proposed Method}
\label{sec:method}


For a user $i$ and an item $j$, the goal of our model estimates the preference score $\mathbf{R}_{ij}$.
As illustrated in Fig.~\ref{fig:overview}, the user representation $\mathbf{u}_i \in \mathbb{R}^D$ is simply learned with an embedding layer, similarly to the traditional collaborative filtering models.
In order to treat cold-start items, however, item representations are learned from their content information.
Given $C$ content information $\{\mathbf{x}^{(1)}_j, ..., \mathbf{x}^{(C)}_j\}$ for each item $j$, where $\mathbf{x}^{(c)}_j \in \mathbb{R}^{D_c}$ for $c = 1, ..., C$ and each $\mathbf{x}^{(c)}_j$ is potentially in different forms from various modalities,
our model feeds each of them into a modality-specific encoder to embed them into a common embedding space. This embedding is annotated by $\mathbf{z}^{(c)}_j \in \mathbb{R}^{d}$ for $c = 1, ..., C$.
If $C > 1$, all $C$ content embeddings are fused into a single item representation $\mathbf{v}_j \in \mathbb{R}^D$ by the Feature Fusion Layer (See Sec.~\ref{sec:method:fusion}).
The final preference ${\mathbf{R}}_{ij}$ is estimated by the dot-product of user and item embeddings; that is, $\hat{\mathbf{R}}_{ij} = {\mathbf{u}_i}^\top \mathbf{v}_j$.

The overall architecture might look standard in recommendation literature; however, we adopt Transformer-based architectures universally for all Modality-specific Encoders, enabling flexible contextualization and fusion across different features.
More details on how to represent each modality will be described in Sec.~\ref{sec:method:modality_specific}.

\subsection{Modality-specific Encoders}
\label{sec:method:modality_specific}

We elaborate our Transformer-based modality-specific encoders, illustrating visual and text representation modules representatively.
We emphasize, however, any feature type can be applied similarly.

\subsubsection{Image Encoder}
\label{sec:method:modality_specific:image}

The left-most box in Fig.~\ref{fig:model_encoders} illustrates our Image Encoder.
To be uncluttered, we omit the feature index superscript $(c)$ and the item index subscript $j$ inside each modality-specific encoder.
Given an image $\mathbf{x} \in \mathbb{R}^{H \times W \times 3}$, where $H$ and $W$ are its height and width, respectively, it is divided into $P \times P$ smaller image patches, forming a set $\{\mathbf{x}_{[1,1]}, \mathbf{x}_{[1,2]}, ..., \mathbf{x}_{[h,w]}\}$, where $h = H/P$, $w = W/P$, and $\mathbf{x}_{[a,b]} \in \mathbb{R}^{P \times P}$ for $a = 1, ..., h$ and $b = 1, ..., w$.
Adopting ViT~\cite{dosovitskiy2020vit}, our Image Encoder first linearly maps the input patches $\{\mathbf{x}_{[1,1]}, \mathbf{x}_{[1,2]}, ..., \mathbf{x}_{[h,w]}\}$ to an embedding space with the Patch Embedding Layer, where the resulting embeddings are denoted by $\{\mathbf{z}_{[1,1]}, \mathbf{z}_{[1,2]}, ..., \mathbf{z}_{[h,w]}\}$, where $\mathbf{z}_{[a,b]} \in \mathbb{R}^d$ for $a = 1, ..., h$ and $b = 1, ..., w$.
Then, following common practice, learnable positional encodings $\{\mathbf{p}_{[a,b]} \in \mathbb{R}^d \}$ are added to the patch embeddings, depending on the location of each patch within the image.
They are fed into $L_c$ Transformer Encoder blocks, where $c = 1, ..., C$ is the content attribute index, contextualizing them by repeated multi-head self-attention and multi-layer perceptrons.
During this process, each patch embedding is updated to capture diverse semantics (\emph{e.g.}, objects and their relations) in the image.
The output is the transformed sequence embeddings $\{\mathbf{z}_{[1,1]}, \mathbf{z}_{[1,2]}, ..., \mathbf{z}_{[h,w]}\}$ from the last Transformer block.
Optionally, an additional \texttt{[CLS]} token is appended to the sequence.
The output embedding corresponding to this \texttt{[CLS]}, denoted by $\mathbf{z} \in \mathbb{R}^d$, encodes semantics of the entire image $\mathbf{x}$.
Either the entire sequence or this aggregated $\mathbf{z}$ is used depending on the feature fusion methods (Sec.~\ref{sec:method:fusion}).

\subsubsection{Video Encoder}
\label{sec:method:modality_specific:video}

The second encoder in Fig.~\ref{fig:model_encoders} illustrates the Video Encoder.
Instead of a single image, it takes as input a video $\mathbf{x} \in \mathbb{R}^{T \times H \times W \times 3}$ with $T$ frames.
Among $T$ video frames, we first randomly sample a clip of $F$ consecutive frames, denoted by \{$\mathbf{x}_{[1]}$, $\mathbf{x}_{[2]}$, ..., $\mathbf{x}_{[F]}$\}.
Then, we adopt a two-stage architecture where we first compute the frame-level embeddings $\{\mathbf{z}_{[1]}, \mathbf{z}_{[2]}, ..., \mathbf{z}_{[F]}\}$, where $\mathbf{z}_{[f]} \in \mathbb{R}^d$ for each frame $f = 1, ..., F$, using the Image Encoder with \texttt{[CLS]} (Sec.~\ref{sec:method:modality_specific:image}).
Then, a learnable temporal positional encoding $\{\mathbf{p}_{[1]}, ..., \mathbf{p}_{[F]} \in \mathbb{R}^d \}$ is added.
The sequence is fed into an additional Transformer Encoder.
While the Image Encoder captures the spatial semantics of each frame, this second-level Transformer Encoder is in charge of capturing temporal semantics in the clip.
Similarly to the image case, a classification token (\texttt{[CLS]}) may be appended to the sequence to aggregate the entire clip representation $\mathbf{z} \in \mathbb{R}^d$ out of it, or the entire output sequence $\{\mathbf{z}_{[1]}, \mathbf{z}_{[2]}, ..., \mathbf{z}_{[F]}\}$ is kept depending on the fusion method.

We choose this two-stage architecture in order to effectively capture the spatio-temporal semantics of the video, including both details at frame level and overall information flow through the temporal axis.
The architecture is similar to the model 2 of the Video Vision Transformer (ViViT)~\cite{arnab2021vivit}, reported as the most efficient and cost-effective.
In order to learn complex underlying spatio-temporal dynamics from videos, choosing a computationally efficient architecture is critically important.

\subsubsection{Text Encoder}
\label{sec:method:modality_specific:text}

Similarly to the visual modalities, we use a Transformer-based Text Encoder similar to BERT~\cite{Devlin2018bert}, illustrated in the right-most box in Fig.~\ref{fig:model_encoders}.

Given a sequence $\{\mathbf{x}_{[1]}, ..., \mathbf{x}_{[T]}\}$ of $T$ words (or sub-words, depending on the particular tokenizer used), where $\mathbf{x}_{[t]} \in \{1, ..., |V|\}$ for $t = 1, ..., T$ and $V$ is the vocabulary set, the Word Embedding Layer encodes them into a sequence of word embeddings, $\{\mathbf{z}_{[1]}, ..., \mathbf{z}_{[T]}\}$ with $\mathbf{z}_{[t]} \in \mathbb{R}^{d}$ for each $t = 1, ..., T$.
Then, they are added with the positional encoding $\{\mathbf{p}_{[1]}, ..., \mathbf{p}_{[T]} \in \mathbb{R}^d\}$.
Unlike the learnable positional encodings used for visual modalities, we follow the fixed positional encodings following BERT~\cite{Devlin2018bert}, as it is more suitable for text.
The position-aware word embeddings pass through a Transformer Encoder which contextualizes the word embeddings throughout the entire text and produces another sequence of transformed word representations.

\subsection{Feature Fusion}
\label{sec:method:fusion}

Once $C$ content information is represented in a common embedding space, $\mathbf{z}_j^{(1)}, ..., \mathbf{z}_j^{(C)} \in \mathbb{R}^d$ for each item $j$, we fuse them into a single item embedding $\mathbf{v}_j \in \mathbb{R}^D$ through the Feature Fusion Layer.
This fusion can be implemented in a variety of ways, described below.

\subsubsection{Late Fusion}
\label{sec:method:fusion:late}

Different content signals are fused at the last step, where each modality-specific encoder provides an aggregated single embedding $\mathbf{z}^{(c)} \in \mathbb{R}^d$ for $c = 1, ..., C$, corresponding to each content signal.
For this, an additional \texttt{[CLS]} token is appended to the input sequence to each modality-specific encoder, illustrated in Fig.~\ref{fig:model_encoders}.
Without being biased to any specific token, the output embedding $\mathbf{z}^{(c)} \in \mathbb{R}^d$ corresponding to this \texttt{[CLS]} token is used to represent each content signal $c$. 
Using the concatenation approach among various fusion methods, we map $\{\mathbf{z}^{(1)}, ...,\mathbf{z}^{(C)}\}$ to the final item representation, $\mathbf{v} \in \mathbb{R}^D$, with a few fully-connected layers in the Feature Fusion Layer; that is, $\mathbb{R}^{C \times d} \rightarrow \mathbb{R}^{D}$.
Thanks to the flexibility of MLPs, $D$ is not necessarily same as $d$.

Note that each feature $\mathbf{z}_j^{(c)}$ may not be aligned yet even though they are in the same embedding space. To further align multimodal information for the same item $j$, we may additionally apply Multimodal Alignment Loss, detailed in Sec.~\ref{sec:method:objective:multimodal}.

\subsubsection{Early Fusion}
\label{sec:method:fusion:early}

In contrast to the Late Fusion where we use only a single output embedding from each modality-specific encoder, 
with Early Fusion, information across different content signals are fused before the token embeddings are aggregated.
That is, we fuse all output tokens $\{\mathbf{z}_{[t]}^{(c)} \in \mathbb{R}^d \}$ from the modality-specific encoders for $t = 1, ..., T_c$ and $c = 1, ..., C$.
An additional Transformer Encoder is adopted for this, taking all tokens $\{\mathbf{z}_{[t]}^{(c)} \in \mathbb{R}^d \}$ to capture dependencies between them by cross-modal attention~\cite{Lu2019vilbert}.
At the end, all contextualized token embeddings may be averaged, or an additional \texttt{[CLS]} token is appended, to aggregate the semantics of the target item, denoted by $\mathbf{v}_j \in \mathbb{R}^D$.
Note that $D = d$ here due to restriction of the Transformer architecture that token embedding size should be always the same. To have $D \ne d$, a fully-connected layer may be added in the end.

With this Early Fusion, additional multimodal alignment may not be necessary, since tokens from all different side information are aligned within the Transformer structure described above.

\subsubsection{Mixture of Late and Early Fusions}

Between the two extreme cases of Late (Sec.~\ref{sec:method:fusion:late}) and Early (Sec.~\ref{sec:method:fusion:early}) Fusions, there are various possibility of mixing those two.
For instance, only a subset of features are fused early and fed into the fusion Transformer.

\subsection{Training Objectives}
\label{sec:method:objective}

The entire model is trained \emph{end-to-end} using the two losses: Rating Ranking Loss and Multimodal Alignment Loss.

\subsubsection{Rating Ranking Loss.}
\label{sec:method:objective:rating}

We train the model to predict higher scores for preferred items and lower scores for the others.
That is, the model is trained to maximize $\{\hat{\mathbf{R}}_{ij}: \mathbf{R}_{ij}=1\}$ and minimize $\{\hat{\mathbf{R}}_{ij}: \mathbf{R}_{ij}=0\}$.
We use contrastive loss, which has been widely adopted for representation learning ~\cite{wang2020understanding,he2020momentum,le2020contrastive,chen2020simclr}.

Specifically, for each user, the item paired in the same example (\emph{i.e.}, this user actually likes the item) is used as positive, while all other items belonging to different pairs in the minibatch are considered as negatives.
With contrastive loss, the encoder is trained to maximize the dot product between the user and item embeddings in the same pair, while minimizing that of the different pairs in the mini-batch.
The Rating Ranking Loss $\mathcal{L}_\text{R}$ for a pair of a user $i$ and an item $j$ is defined as
\begin{align}
   \mathcal{L}_\text{R} = 
   & - \log \frac{ \exp ({\mathbf{u}_{i}}^\top \mathbf{v}_{j})}{ \sum_{j' \in \mathcal{B}} \exp ({\mathbf{u}_{i}}^\top \mathbf{v}_{j'})} \nonumber \\
   & - \log \frac{ \exp ({\mathbf{u}_{i}}^\top \mathbf{v}_{j})}{ \sum_{i' \in \mathcal{B}} \exp ({\mathbf{u}_{i'}}^\top \mathbf{v}_{j})},
  \label{eq:loss_rr}
\end{align}
where $\mathcal{B}$ is the set of user-item pairs in the minibatch.

Here, one might argue that the items in the minibatch other than the paired one with the user might not be actually negative. Unlike classification models like SimCLR~\cite{chen2020simclr}, the user might actually like additional items other than the currently paired one.
We thus optionally filter out these false negatives from the denominator of Eq.~\eqref{eq:loss_rr} for more precise training. We report empirical performance with or without false negative filtering in Sec.~\ref{sec:exp:ablation}.

\subsubsection{Multimodal Alignment Loss}
\label{sec:method:objective:multimodal}

With the Late Fusion of multiple ($C > 1$) content features, an additional Multimodal Alignment Loss can be beneficial, as various content features may not be aligned yet in the common embedding space.
Specifically, we apply contrastive loss to all item embeddings within the minibatch, maximizing the similarity between content embeddings for the same item, while minimizing it between all other combinations. Multimodal Alignment Loss $\mathcal{L}_\text{M}$
is defined by
\begin{align}
  \mathcal{L}_\text{M} = 
    - \log \frac{1}{Z} \sum_{c=1}^C \sum_{c' > c}  e^{{\mathbf{z}_j^{(c)}}^\top \mathbf{z}_j^{(c')}},\\
    \text{with }
    Z = \sum_{j' \in \mathcal{B}} \sum_{c=1}^C \sum_{c'=1}^C e^{{\mathbf{z}_j^{(c)}}^\top \mathbf{z}_{j'}^{(c')}},
    \nonumber
\end{align}
where $\mathcal{B}$ is the set of items in the minibatch.
When $\mathcal{L}_\text{M}$ is used, we linearly combine it with $\mathcal{L}_\text{R}$; that is,
\begin{equation}
  \mathcal{L} = \mathcal{L}_\text{R} + \lambda \mathcal{L}_\text{M},    
\end{equation}
where $\lambda$ controls relative importance of the two losses.

\subsection{Inference}

For videos, recall that we randomly sample a segment with $F$ frames at training.
At inference, we sample $S > 1$ segments and predict preference scores with each of them. Then, we aggregate those scores by taking the max:
\begin{equation}
    \hat{\mathbf{R}}_{ij} = \max_{s = 1, ..., S} {\mathbf{u}_i}^\top \mathbf{v}_{j_s},
\end{equation}
where $\mathbf{v}_{j_s}$ is the video embedding based on the segment $s$ for the item $j$.
In this way, we cover wider range of the video and compute the score based on a segment that the user most likely prefers.

For the news domain, the content representation embedding is directly generated by feeding the entire image and text without any cropping, so we simply estimate by
\begin{equation}
  \hat{\mathbf{R}}_{ij} = {\mathbf{u}_i}^\top \mathbf{v}_j.
\end{equation}

\section{Experiments}
\label{sec:exp}

We conduct extensive experiments to verify the effectiveness of our framework on multiple recommendation domains and datasets.

\subsection{Experimental Settings}



\begin{table*}
  \caption{Overview of Our Datasets}
  \centering
  { 
  \begin{tabular}{ l|rrrr|c }
    \toprule
    Dataset & Users & Items & Ratings & Density & Domain \\
    \midrule
    MovieLens 25M & 162,541 & 62,423 & 25,000,095 & 0.246\% & Movie \\
    Yahoo Movies & 7,642 & 11,915 & 211,231 & 0.232\% & Movie \\
    Chosun News 2022 & 389,188 & 288,540 & 42,457,502 & 0.038\% & News \\
    \bottomrule
  \end{tabular}}
  \label{tab:dataset}
\end{table*}

\vspace{0.1cm}
\emph{Datasets.}
We choose the movie and news domain for our experiments. The movie domain is chosen as it contains the richest content signals, \emph{e.g.}, visual scenes, textual summary or script, metadata like genre, director, or main actors. We choose the news domain due to its cold-start nature; that is, recent news articles are mostly valuable to recommend.
As listed in Table~\ref{tab:dataset}, we use two widely-used standard benchmarks on the movie domain, MovieLens 25M~\cite{harper2015movielens} and Yahoo Movies~\cite{sahoo2008multi}.
For the news domain, we use Chosun News 2022, containing all the articles and user activities between January and December 2022 on \texttt{chosun.com}, one of the most representative newspapers in Korea.
Both MovieLens and Yahoo Movies provide explicit ratings from 1 (least preferred) to 5 (most preferred), so we convert them to implicit ones with 3.5 as the threshold, following ~\cite{Lee2018cdml,zhou2018deep}.
Chosun News dataset considers a click from a user on a news article as a positive feedback, and negative otherwise.

We exclude items with any missing content information from all datasets. Also, we filter out users with less than 20 ratings from MovieLens, following~\cite{weimer2007cofirank,lee2014lcr}. We do not filter out ratings from Yahoo Movies and Chosun News.
After filtering, we randomly split the items into training, cold validation, and cold test with the ratio of 85:7.5:7.5 for MovieLens and 70:15:15 for Yahoo Movies. For Chosun News, we use activities from the first 6 months for training, next 3 months for validation, and the rest for testing.
The cold validation set is used to tune hyper-parameters, and the cold test set is used to evaluate the final performance.
The two movie datasets contain only 891 overlapping movies, $\sim$2.77\% out of 32,156 movies in total (regarding the transfer learning experiment in Sec.~\ref{sec:exp:representation}).


\vspace{0.1cm}
\emph{Content Features.}
As all three datasets provide limited content signals, we collect additional visual and text data.

For visual content of the movie datasets, we use movie trailers provided by MovieLens~\cite{movielens_trailer} and MovieNet~\cite{huang2020movienet}, since the full videos are publicly unavailable for most movies due to copyright.
From each video, frames of size $224 \times 224$ are sampled at 2 fps. We drop the first and last 10\% of the sampled frames, since they are often age rating screen or ending credits.
The average length of the trailers is 137 seconds for MovieLens and 140 seconds for Yahoo Movies, so we get around 220 frames per video on average for both datasets.
For visual content of the news dataset, we use up to 3 images collected from the web queried by the title of each article.

For text content of the movie datasets, we use movie synopsis collected from \texttt{imdb.com} for MovieLens. Yahoo Movies self-contains synopsis. These synopses are 2--3 sentences that summarize the movie overview.
The sentences are first tokenized at word level with the maximum length of 512, using uncased $\text{BERT}_\text{BASE}$ tokenizer~\cite{Devlin2018bert} with $|V| = 30,522$. 
The average number of text tokens is $54.7$ and $83.0$ for MovieLens and Yahoo Movies, respectively.
For Chosun News, we use the title and the body text of the article, following the same preprocessing above. The average number of tokens in this dataset is 257.3.
We use the $\text{KoBERT}_\text{Base-V1}$ tokenizer~\cite{jeon2019korean} with $|V| = 8,002$ for Korean language in Chosun News.

\vspace{0.1cm}
\emph{Evaluation Metrics.}
We measure recommendation performance by ranking all unseen items for each user in a held-out test set and comparing the top $K$ items from the ranked list with the items that the user actually gave positive feedback to.
Following CLCRec~\cite{wei2021clcrec}, we treat all users with varied number of ratings equally by averaging the score for each user.
We use three widely-used metrics for ranking tasks: \{Precision, Recall, NDCG\}@$K$ with $K = \{1, 5, 10, 20\}$.

\vspace{0.1cm}
\emph{Competing Models.}
We compare with 4 recent cold-start item recommendation models using content information: CLCRec~\cite{wei2021clcrec}, DropoutNet~\cite{volkovs2017dropoutnet}, CVAR~\cite{zhao2022cvar}, and PMGT~\cite{liu2021pmgt}.
For fair comparison, we use the same set of multimedia features for all models, not the categorical side information used in baseline papers (\emph{e.g.}, \cite{zhao2022cvar}); that is,
ViT~\cite{dosovitskiy2020vit} embeddings pretrained on ImageNet~\cite{deng2009imagenet} and BERT~\cite{Devlin2018bert} embeddings for visual and text features, respectively.

\vspace{0.1cm}
\emph{Model Hyperparameters.}
We experiment with $D = $ \{32, 64, 128, 256, 512, 1024, 2048\} for the user ($\mathbf{u}_i$) and item ($\mathbf{v}_j$) embedding size.
The token embedding size $d$ is set to 192, following \cite{arnab2021vivit}.
For visual features, we spatially split each frame to $P \times P$ patches with $P = 16$.
We stack $L = 4$ Transformer blocks for the Image and Video Encoders, and $L = 12$ for the Text Encoder.
Positional encodings $\mathbf{p}$ for visual encoders are learned from data~\cite{dosovitskiy2020vit}.
Visual encoders are trained from scratch to avoid use of classification labels, while the text encoder starts from the pretrained BERT~\cite{Devlin2018bert}.
For the Feature Fusion Layer, we try Late Fusion with a single or two-layer MLP and Early Fusion with a single Transformer layer (Sec.~\ref{sec:exp:ablation}).
We perform grid search for $\lambda$ within the range $[0, 1]$. We randomly sample $S=10$ segments for video inference.

\begin{table*}
  \caption{Comparison with the Baselines on All Datasets (\%)}
  \centering
  { 
  \begin{tabular}{ l | l | rrrr | rrrr | rrrr }
    \toprule
    \multirow{2}{*}{Dataset} & \multirow{2}{*}{Method} & \multicolumn{4}{c|}{NDCG ($\uparrow$)} & \multicolumn{4}{c|}{Precision ($\uparrow$)} & \multicolumn{4}{c}{Recall ($\uparrow$)} \\
    &  & @1 & @5 & @10 & @20 & @1 & @5 & @10 & @20 & @1 & @5 & @10 & @20 \\
    \midrule
    \multirow{5}{*}{MovieLens} 
    & {DropoutNet~\cite{volkovs2017dropoutnet}} 
    & 7.33 & 4.99 & 4.72 & 5.29     & 7.33 & 4.36 & 4.27 & 4.79     & 7.33 & 4.38 & 4.34 & 5.54 \\
    & {CLCRec~\cite{wei2021clcrec}}
    & 9.09 & 6.59 & 6.77 & 8.62  & 9.09 & 6.02 & 6.31 & 7.19 & 9.09 & 6.06 & 6.71 & 10.27 \\
    & {CVAR~\cite{zhao2022cvar}} 
    & 8.89 & 9.11 & 9.09 & 9.49     & 8.89 & 9.12 & 9.10 & \textbf{9.56}     & 8.89 & 9.13 & 9.12 & 9.71 \\
    & {PMGT~\cite{liu2021pmgt}} 
    & \textbf{14.13} & 7.64 & 7.22 & 8.54 & \textbf{14.13} & 6.20 & 6.24 & 7.54 & \textbf{14.13} & 6.22 & 6.43 & 8.98\\
    & {Ours} & 14.05 & \textbf{11.41} & \textbf{10.16} & \textbf{10.40} & 14.05 & \textbf{10.77} & \textbf{9.13} & 8.14 & 14.05 & \textbf{10.80} & \textbf{9.39} & \textbf{11.33} \\
    \midrule
    \multirow{5}{*}{Yahoo Movies}
    & {DropoutNet~\cite{volkovs2017dropoutnet}} 
    & 1.10 & 1.68 & 2.40 & 3.29 & 1.10 & 1.12 & 1.10 & 1.01 & 1.10 & 2.16 & 4.05 & 6.40 \\
    & {CLCRec~\cite{wei2021clcrec}} 
    & 0.75 & 6.50 & 6.47 & 6.65 & 0.75 & 3.87 & 2.24 & 1.24 & 0.75 & 7.95 & 8.34 & 8.97 \\
    & {CVAR~\cite{zhao2022cvar}} 
    & 1.07 & 1.74 & 2.31 & 3.09 & 1.07 & 1.34 & 1.17 & 1.05 & 1.07 & 2.67 & 4.13 & 6.49 \\
    & {PMGT~\cite{liu2021pmgt}}
    & 2.04 & 4.55 & 5.27 & 6.47 & 2.04 & 3.38 & 2.33 & 1.88 & 2.04 & 6.10 & 8.48 & 12.31 \\
    & {Ours} 
    & \textbf{5.39} & \textbf{8.53} & \textbf{12.42} & \textbf{12.48} & \textbf{5.39} & \textbf{5.87} & \textbf{5.74} & \textbf{6.27} & \textbf{5.39} & \textbf{8.86} & \textbf{15.14} & \textbf{16.24} \\
    \midrule
    \multirow{5}{*}{Chosun News}
    & {DropoutNet~\cite{volkovs2017dropoutnet}} 
    & 2.43 & 1.92 & 2.18 & 2.17 & 2.43 & 2.07 & 2.07 & 2.14 & 2.43 & 2.08 & 2.11 & 2.22 \\
    & {CLCRec~\cite{wei2021clcrec}} 
    & 1.52 & 1.36 & 1.29 & 1.23 & 1.52 & 1.40 & 1.28 & 1.19 & 1.52 & 1.40 & 1.30 & 1.23 \\
    & {CVAR~\cite{zhao2022cvar}} 
    & 3.34 & 3.11 & 2.75 & 2.63 & 3.34 & 2.92 & 2.61 & 2.34 & 3.34 & 2.93 & 2.66 & 2.41 \\
    & {PMGT~\cite{liu2021pmgt}} 
    & 2.80 & 2.50 & 2.19 & 2.20 & 2.80 & 2.30 & 1.93 & 2.00 & 2.80 & 2.30 & 1.93 & 2.10\\
    & {Ours} 
    & \textbf{4.13} & \textbf{4.06} & \textbf{3.52} & \textbf{3.68} & \textbf{4.13} & \textbf{4.13} & \textbf{3.49} & \textbf{3.47} & \textbf{4.13} & \textbf{4.14} & \textbf{3.50} & \textbf{3.53} \\
    \bottomrule
  \end{tabular}}
  \label{tab:baselines}
\end{table*}

\vspace{0.1cm}
\emph{Training Hyperparameters.}
We use Adam optimizer~\cite{kingma2014adam} with $\beta_1$ = 0.9, $\beta_2$ = 0.999, and $\epsilon$ = $10^{-8}$. We linearly warm up the learning rate during the first 3 epochs, and train up to 200 epochs. After 70\% of training, we decay the learning rate to 20\% of the initial one, which is found by grid search among $\{10^{-6}, 10^{-5}, 10^{-4}, 10^{-3}, 10^{-2}\}$.
We use batch size $B = 48$. For the movie datasets, a single sub-clip of length $F = 32$ is randomly sampled within each trailer, allowing the model to see various parts of the video uniformly throughout the whole training process.
All reported results are averaged over five experiments with random initialization.



\subsection{Comparison to the Baselines}

\begin{table*}
  \centering
  \caption{Evaluation on General (Warm + Cold) Recommendation Task (YM)}
  \label{tab:warm_cold}
  { 
  \begin{tabular}{ l | rrrr | rrrr | rrrr }
    \toprule
    \multirow{2}{*}{Modality} & \multicolumn{4}{c|}{NDCG ($\uparrow$)} & \multicolumn{4}{c|}{Precision ($\uparrow$)} & \multicolumn{4}{c}{Recall ($\uparrow$)} \\
    & @1 & @5 & @10 & @20 & @1 & @5 & @10 & @20 & @1 & @5 & @10 & @20 \\
    \midrule
    {DropoutNet~\cite{volkovs2017dropoutnet}} 
    & 2.04 & 1.91 & 2.41 & 2.59 & 2.04 & 1.53 & 1.52 & 1.09 & 2.04 & 1.99 & 3.25 & 3.79 \\
    {CVAR~\cite{zhao2022cvar}} 
    & 2.31 & 1.98 & 2.35 & 2.62 & 2.31 & 1.66 & 1.47 & 1.15 & 2.31 & 2.14 & 3.18 & 3.89 \\
    Ours
    & \bf{6.71} & \bf{5.94} & \bf{6.33} & \bf{6.94} & \bf{6.71} & \bf{3.75} & \bf{2.63} & \bf{1.70} & \bf{6.71} & \bf{6.56} & \bf{7.73} & \bf{9.59} \\
    \bottomrule
  \end{tabular}}
\end{table*}

\begin{table*}
  \centering
  \caption{Modality Ablation Study (CN)}
  \label{tab:modality_ablation}
  \centering
  { 
  \begin{tabular}{ l | rrrr | rrrr | rrrr }
    \toprule
    \multirow{2}{*}{Modality} & \multicolumn{4}{c|}{NDCG ($\uparrow$)} & \multicolumn{4}{c|}{Precision ($\uparrow$)} & \multicolumn{4}{c}{Recall ($\uparrow$)} \\
    & @1 & @5 & @10 & @20 & @1 & @5 & @10 & @20 & @1 & @5 & @10 & @20 \\
    \midrule
    Visual Only
    & 0.61 & 1.06 & 1.02 & 1.01 & 0.61 & 1.22 & 1.07 & 0.93 & 0.61 & 1.22 & 1.07 & 1.27 \\
    Text Only
    & 3.51 & 2.69 & 2.87 & 2.83 & 3.51 & 2.76 & 2.97 & 2.86 & 3.51 & 2.76 & 2.97 & 2.88 \\
    Visual + Text
    & \bf{4.13} & \bf{4.06} & \bf{3.52} & \bf{3.68} & \bf{4.13} & \bf{4.13} & \bf{3.49} & \bf{3.47} & \bf{4.13} & \bf{4.14} & \bf{3.50} & \bf{3.53} \\
    \bottomrule
  \end{tabular}}
\end{table*}
  
\begin{table*}
  \centering
  \caption{MLP Architecture (CN)}
  \label{tab:mlp_structure}
  { 
  \begin{tabular}{ l | rrrr | rrrr | rrrr }
    \toprule
    \multirow{2}{*}{Modality} & \multicolumn{4}{c|}{NDCG ($\uparrow$)} & \multicolumn{4}{c|}{Precision ($\uparrow$)} & \multicolumn{4}{c}{Recall ($\uparrow$)} \\
    & @1 & @5 & @10 & @20 & @1 & @5 & @10 & @20 & @1 & @5 & @10 & @20 \\
    \midrule
    No FC layer
    & 3.72 & 3.41 & 3.40 & 3.38 & 3.72 & 3.53 & 3.44 & 3.26 & 3.72 & 3.53 & 3.44 & 3.56 \\
    1 FC layer
    & \bf{4.32} & 3.72 & 3.52 & 3.66 & \bf{4.32} & \bf{3.84} & 3.50 & \bf{3.59} & \bf{4.32} & \bf{3.84} & 3.52 & 3.62 \\
    2 FC layers
    & 4.02 & \bf{3.89} & \bf{3.77} & \bf{3.68} & 4.02 & \bf{3.84} & \bf{3.69} & 3.58 & 4.02 & \bf{3.84} & \bf{3.69} & \bf{3.63} \\
    \bottomrule
  \end{tabular}}
\end{table*}
  
\begin{table*}
  \centering
  \begin{minipage}{.45\linewidth}
    \centering
    \caption{Feature Fusion (YM, CN)}
    \label{tab:feature_fusion}
    { 
    \begin{tabular}{ cl | rrr }
      \toprule
      Data & Fusion 
      & NDCG@10 & Prec@10 & Recall@10 \\
      \midrule
      YM & Late  
      & \textbf{12.42} & \textbf{5.74} & \textbf{15.14} \\
         & Early 
         & 10.00 & 4.19 & 14.18 \\
      \midrule
      CN & Late  
      & 3.40 & 3.43 & 3.44 \\
         & Early 
         & \bf{3.52} & \bf{3.49} & \bf{3.50} \\
      \bottomrule
    \end{tabular}}
  \end{minipage}
  \begin{minipage}{.4\linewidth}
    \centering
    \caption{Embedding Size (YM)}
    \label{tab:ablation_architecture}
    { 
    \begin{tabular}{ c | rrr }
      \toprule
      $D$ & NDCG@10 & Prec@10 & Recall@10 \\
      \midrule
      32   & 8.87 & 4.10 & 9.72 \\
      64   & 10.21 & 4.55 & 11.48 \\
      128  & 11.37 & 5.59 & 14.17 \\
      256  & 11.84 & 5.61 & 14.16 \\
      512   & 12.02 & \bf{5.79} & 14.46 \\
      1024  & \bf{12.42} & 5.74 & \bf{15.14} \\
      2048  & 12.17 & 5.66 & 14.56 \\
      \bottomrule
    \end{tabular} \vspace{0.5cm}}
    %
  \end{minipage}
  \begin{minipage}{.33\linewidth}
    \centering
    \caption{False Negative Filter (ML)}
    \label{tab:false_negative}
    {\footnotesize 
    \begin{tabular}{ c | rrr }
      \toprule
      Filtering & NDCG@10 & Prec@10 & Recall@10 \\
      \midrule
      Yes  & 9.85 & \textbf{9.14} & 9.21 \\
      No   & \textbf{10.16} & 9.13 & \textbf{9.39} \\
      \bottomrule
    \end{tabular}}
  \end{minipage}
  \vspace{0.5cm}
  \\ \small{ 
    *Datasets: ML = MovieLens, ~YM = Yahoo Movies, 
    CN = Chosun News}
\end{table*}

Table~\ref{tab:baselines} reports the performance on cold-start recommendation evaluated by \{NDCG, Prec, Recall\}@$K$ with $K = \{1,5,10,20\}$.
Our model outperforms all baselines under almost all metrics.
Another notable observation is the relationship between the models' performance and the value of $K$.
On MovieLens, the average number of positive items in the test set is 23.1. 
Our approach tends to be stronger with smaller $K$, so it will be more suitable for cases like watch next, where only the top one item will be auto-played.
Baseline models like CVAR, on the other hand, tend to be stronger with larger $K$, so they will be more suitable for homepage recommendations, where multiple items are presented at the same time.
On Yahoo Movies, the average number of positive items is 3.5,
much lower than 20. Thus, all methods tend to show higher scores with larger $K$.

\subsection{General (Warm) Recommendations}
\label{sec:exp:warm}

Although we focus on cold-start recommendation problem in this paper, we also evaluate on the general recommendation task, where some test items are not necessarily cold-start.
We conduct this experiment on Yahoo Movies, where the training set is the same but the test set consists of both warm and cold items.
We compare our model to DropoutNet and CVAR in Table~\ref{tab:warm_cold}.
As seen in the table, our method consistently outperforms the two strongest baselines not just on the cold-start, but on the general recommendation task.

\subsection{Ablation Study}
\label{sec:exp:ablation}

\textbf{Modality Ablation.}
To explore the effectiveness of multimodal features and their alignment, we compare the performance of our full model against the same model with either visual or text content only.
As seen in Table~\ref{tab:modality_ablation}, using multimodal features and alignment loss improve the performance over single-modality baselines.

\vspace{0.1cm} \noindent
\textbf{Model Architecture Ablation.}
We compare the Late and Early Fusions (Sec.~\ref{sec:method:fusion}).
Table~\ref{tab:feature_fusion} reports that the Late Fusion outperforms on YahooMovies, while the Early Fusion slightly performs better on Chosun News.
We conjecture that this difference comes from the nature of each dataset.
Specifically, when two modalities provide more distinct information about the item, referring to tokens from each other is more beneficial during self-attention. The news domain seems more like this, where text tends to provide a chronological outline and interpretation of an incident in abstract, while images provide a snapshot of the event with visual details.
Comparing against baselines in Table~\ref{tab:baselines}, however, our method still outperforms all baselines regardless of the fusion method.

We additionally compare the performance with various number of MLP layers after the fusion.
From the model design perspective, having at least one FC layer is beneficial, since it allows us to arbitrarily set the output embedding dimensionality.
Before fusion, we have $C$ modalities, and simply concatenating the features from each modality results in a $D' = \sum_{c=1}^C d_c$ dimensional vector, where $d_c$ is the feature dimensionality of modality $c$.
Without any FC layer on top of this, this $D'$ becomes the output vector size.
With an FC layer, we can map $D'$ to an arbitrary size $D$.
Having additional FC layers, at least up to 2, is indeed beneficial, as shown in Table~\ref{tab:mlp_structure}.
Stacking more layers shows marginal performance gain, indicating that the complexity of content representations is learned well enough at the lower-level encoders, so the MLP layers can be concise.

\vspace{0.1cm} \noindent
\textbf{Embedding Size Exploration.}
Table~\ref{tab:ablation_architecture} summarizes the performance of our model with different embedding sizes ($D$)
on Yahoo Movies.
As expected, larger embedding size leads to better performance in general.
It peaks around $D = 1024$ and saturates with diminishing returns.
We also observe that $D = 128$ is a good trade-off between the cost and performance, aligned with a previous observation~\cite{Lee2018cdml}.
We use $D = 128$ for other ablation studies for efficient exploration.

\vspace{0.1cm} \noindent
\textbf{False Negatives Filtering.}
To quantify the effect of false negatives discussed in Sec.~\ref{sec:method:objective:rating}, we compare our models with and without false negatives filtering in $\mathcal{L}_\text{R}$ on MovieLens.
Table~\ref{tab:false_negative} shows that this filtering has minimal impact.
We conjecture that false negatives are less likely to be included in a small minibatch as the scale of the dataset gets larger.
Considering additional computational overhead, we conduct all other experiments without it.

\subsection{Content Representation Evaluation}
\label{sec:exp:representation}

\begin{table}
  \caption{Experimental Result on Content Representations}
  \centering
  {
  \begin{tabular}{ ll | rrr }
    \toprule
    Pretraining & Target & N@10 & P@10 & R@10 \\
    \midrule
    From scratch & \multirow{2}{*}{MovieLens} & \textbf{9.50} & \textbf{8.80} &\textbf{9.09}\\
    ViT (ImageNet) & & 5.32 & 5.31 & 5.50 \\
    \midrule
    From scratch & \multirow{3}{*}{Yahoo Movies} & \textbf{7.25} & \textbf{3.63} & \textbf{9.24} \\
    ViT (ImageNet) &  & 1.59 & 0.79 & 2.66 \\
    Ours (MovieLens) &  & \textbf{5.22} & \textbf{2.63} & \textbf{6.90} \\
    \bottomrule
  \end{tabular}}
  *Metrics: N = NDCG, ~P = Precision, ~R = Recall
  \label{tab:transfer}
\end{table}

To verify if our content embeddings properly capture users' watch behavior in general, we conduct two studies of transfer learning.

First, we compare our full model against the same model where the feature extractor is replaced with ViT~\cite{dosovitskiy2020vit} trained on ImageNet, average-pooled over the temporal axis.
Comparing the first four rows in Table~\ref{tab:transfer} reveals the difference of training directly on the user activities \emph{vs.} classification data.
We observe that the performance of our model trained from scratch outperforms the same model using ViT pre-trained embeddings on both MovieLens and Yahoo Movies.
Our hypothesis that classification labels are not the best signal to train on for recommendation purpose is quantitatively confirmed from this result.
From this, we confirm the importance of direct training on recommendation signals, instead of relying on labels for a proxy classification task.

Next, we evaluate transferability of our learned content representation from one dataset to another, to see if the learned content model is general enough to be competent on different set of users.
The last row in Table~\ref{tab:transfer} shows reasonable performance of cold-start recommendation on Yahoo Movies, using content embeddings trained on MovieLens.
Considering the low overlapping movies ($\sim$2.77\%) between these two datasets, our model turns out to truly map the raw content signals to users' taste, successfully transferring user behaviors from one dataset to another.

\subsection{Qualitative Analysis}

\begin{figure*}
  \centering
  \includegraphics[width=0.8\linewidth]{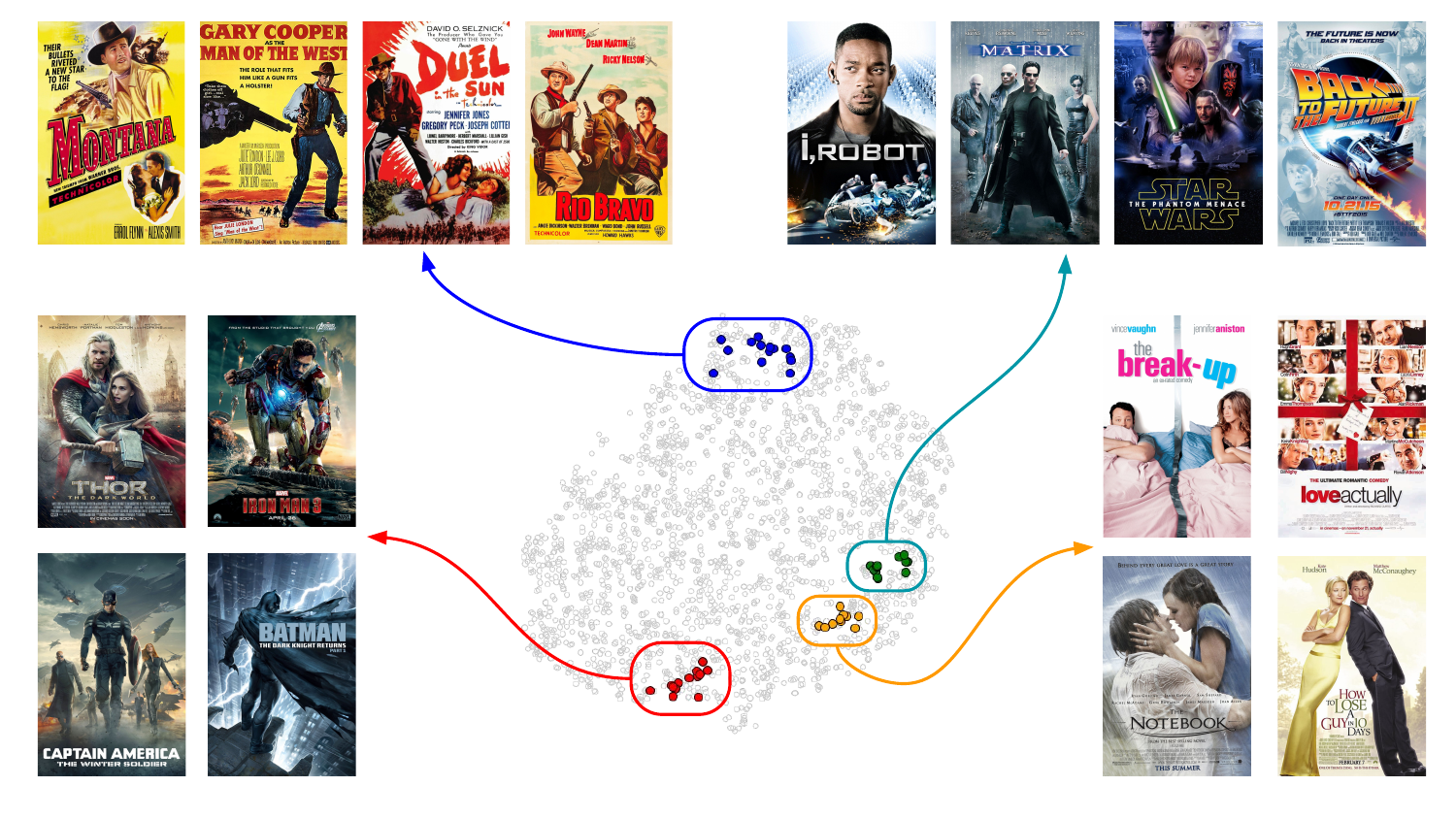}
  \caption{t-SNE Visualization of Learned Video Embeddings}
  \label{fig:tsne}
\end{figure*}

\begin{table*}
  \caption{Full list of the movie titles in the colored clusters in Fig.~\ref{fig:tsne}}
  \centering
  {
  \begin{tabular}{ c | c }
    \toprule
    Red Cluster & Orange Cluster \\
    \midrule
     Batman: The Dark Knight Returns, Part 1 (2012) &  Love Actually (2003) \\
     Batman: Year One (2011) & Break-Up, The (2006) \\
     Superman Unbound (2013) &  Notebook, The (2004) \\
     Captain America: The First Avenger (2011) & How to Lose a Guy in 10 Days (2003) \\
     Captain America: The Winter Soldier (2014) & 12 Dates of Christmas (2011) \\
     Iron Man 2 (2010) & Princess Diaries 2: Royal Engagement, The (2004) \\
     Iron Man 3 (2013) & P.S. I Love You (2007) \\
     Thor: The Dark World (2013) &  Elizabethtown (2005) \\
     Guardians of the Galaxy (2014) & Bridget Jones: The Edge of Reason (2004) \\
     Fantastic Four (2005) & Catch and Release (2006) \\
    \midrule
    Green Cluster & Blue Cluster \\
    \midrule
    I, Robot (2004) & Duel in the Sun (1946) \\
    Star Trek VI: The Undiscovered Country (1991) & Ride Lonesome (1959) \\
    Star Wars: Episode I - The Phantom Menace (1999) & Montana (1950) \\
    Star Wars: Episode II - Attack of the Clones (2002) & Man of the West (1958) \\
    Back to the Future Part II (1989) &  Man Who Never Was, The (1956) \\
    Back to the Future Part III (1990) & Unforgiven, The (1960) \\
    Matrix, The (1999) & Bonnie and Clyde (1967) \\
    Matrix Revolutions, The (2003) & She Wore a Yellow Ribbon (1949) \\
    Battlefield Earth (2000) & Rio Bravo (1959) \\
    Pitch Black (2000) & Hombre (1967) \\
    \bottomrule
  \end{tabular}}
  \label{tab:tsne_full}
\end{table*}

\begin{figure*}
  \centering
  \includegraphics[width=0.85\linewidth]{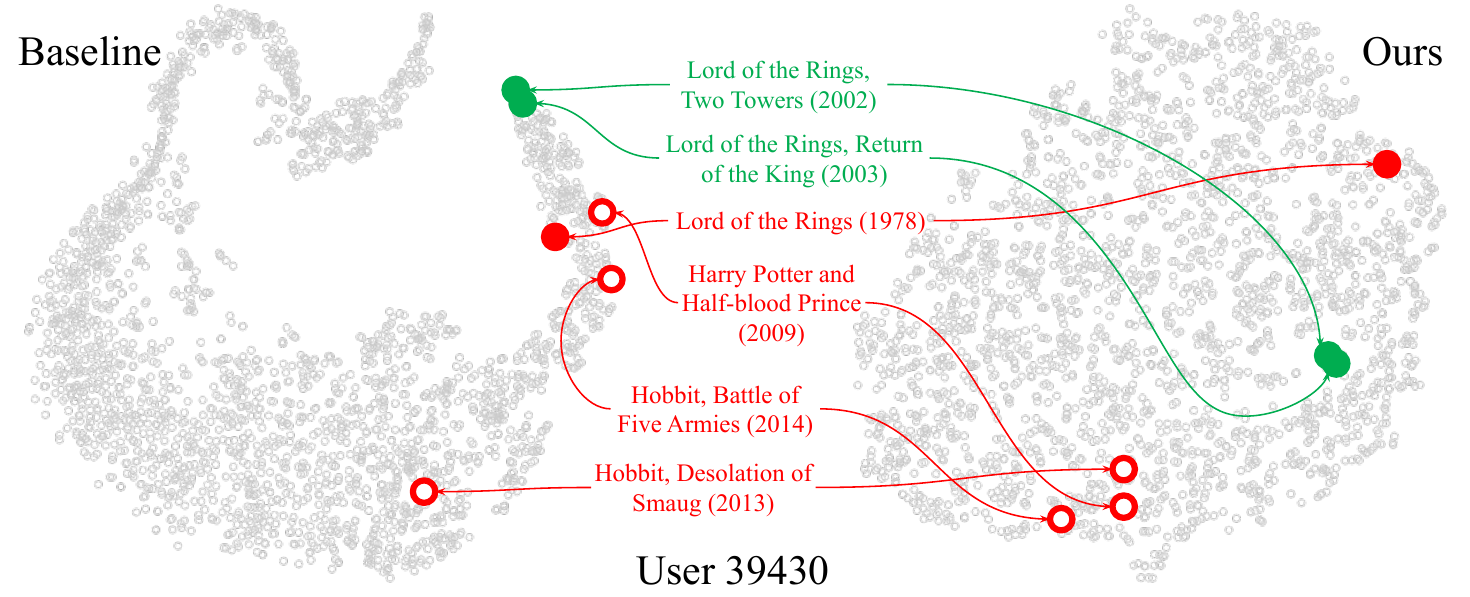}\\
  \includegraphics[width=0.85\linewidth]{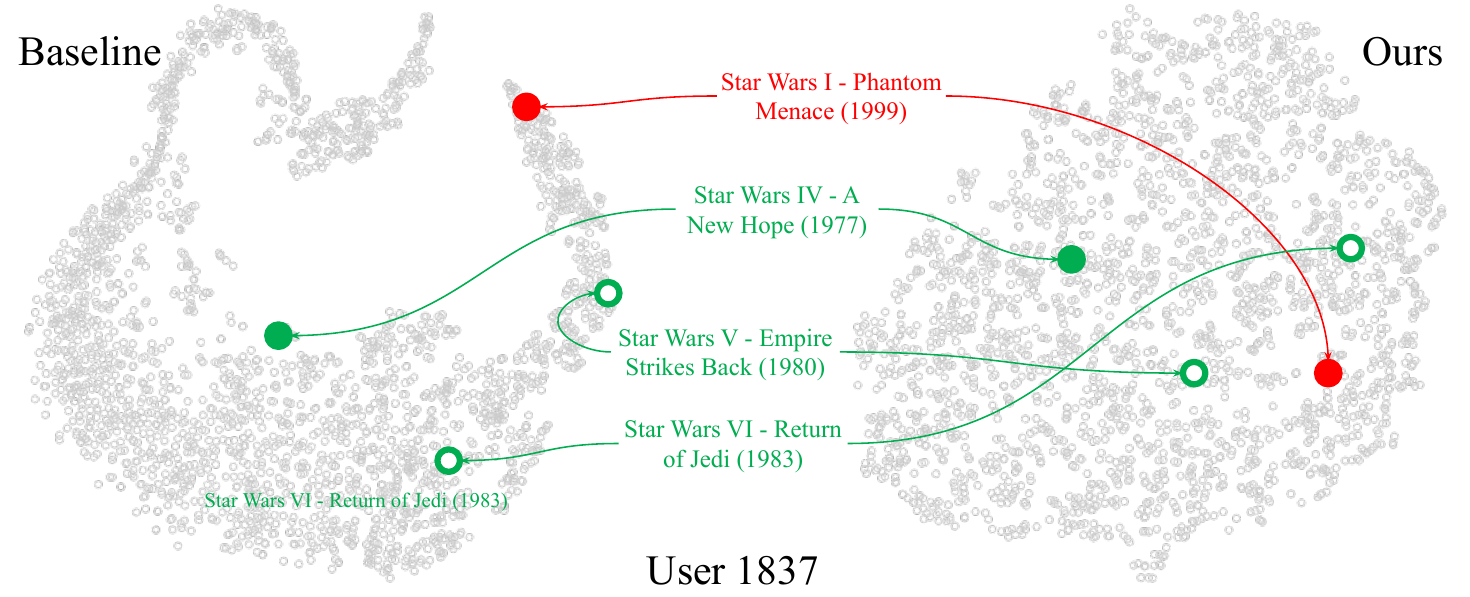}
  \caption{Illustration of item embeddings used in the pairwise similarity analysis in Table~\ref{tab:example_users2}.}
  \label{fig:cossim_exapmles}
\end{figure*}

We visualize the learned video embeddings in 2D for qualitative understanding. 
Fig.~\ref{fig:tsne} presents the t-SNE plot~\cite{van2008visualizing} of the video embeddings learned by our model using visual and text content on MovieLens.
We observe that similar movies are positioned nearby each other in the embedding space. For instance, Fig.~\ref{fig:tsne} illustrates 4 clusters with highly relevant movies in different colors: heroes (red), romantic comedies from mid-2000s (orange), science fictions (green), and western movies from mid-1900s (blue).
The full list of colored dots is listed in Table~\ref{tab:tsne_full}.

\begin{table*}
  \caption{Examples of Fine-grained Taste Estimation. Values in the range [-1, 1] represent preference.}
  \centering
  {
  \begin{tabular}{ lll | crr }
    \toprule
    User & Split & Movie Title & GT & Ours & Baseline \\
    \midrule
    5649 & Train & Lion King (1994) & Like & -- & -- \\
    & & Grease (1978) & Like  & -- & -- \\
    \cmidrule{2-6}
    & Test & Aladdin (1992) & Like & 0.848 & 0.006 \\
    & & Tarzan (1999) & Like & 0.762 & 0.102 \\
    & & Sixth Sense (1999) & Like & 0.424 & -0.056 \\
    & & Sinbad (2003) & Dislike & -0.799 & 0.270 \\

    \midrule

    1837 & Train & Star Wars IV - A New Hope (1977)  & Like & -- & -- \\
    & & Pulp Fiction (1994) & Like & -- & -- \\
    & & Forrest Gump (1994) & Like & -- & -- \\
    \cmidrule{2-6}
    & Test & Star Wars VI - Return of the Jedi (1983) & Like & 0.489 & 0.242 \\
    & & Star Wars V - Empire Strikes Back (1980) & Like & 0.014 & -0.294 \\
    & & Star Wars I - Phantom Menace (1999) & Dislike & -0.276 & 0.462 \\
    \bottomrule
  \end{tabular}}
  \label{tab:example_users}
\end{table*}

For a deeper understanding of the improved performance of our model, we look into actual predicted scores for a couple of users in Table~\ref{tab:example_users}, comparing with the strongest baseline on MovieLens, CLCRec~\cite{wei2021clcrec}.
User 5649 is known to like the Lion King (1994) in the training set. Although this user likes animations, the test set indicates she prefers Aladdin (1992) and Tarzan (1999) (all from Disney), but not Sinbad (2003) from DreamWorks.
Our model captures the user's taste precisely, estimating higher scores, 0.848 and 0.762, for the two preferred items, while significantly lower one (-0.799) for Sinbad.
The baseline, on the other hand, predicts similar scores for all three animations, even slightly higher (0.270) for Sinbad.
This example illustrates that the proposed method trained directly on the user activities is better capable of capturing fine-grained tastes of users than previous works trained on classification labels.

Another example is user 1837. This user likes Star Wars Episode IV, V, and VI, but for some reason not the Episode I. Given the user likes the Episode IV only in the training set, our method retrieves Episodes V and VI (100\% correct), while the baseline model recommends Episode I and VI (50\% correct). Again, this example indicates our approach better captures subtle difference among multiple episodes of the same series, Star Wars, than existing methods.

\begin{table*}
  \caption{Cosine Similarity Examples of Item Embeddings}
  \centering
  {
  \begin{tabular}{ ll | rr }
    \toprule
    User & Comparison & Ours & Baseline \\
    \midrule
    39430 & Like (All \green{$\bullet$}) \emph{vs.} Dislike (All \red{$\bullet \circ$}) & -0.079 & 0.980 \\
    & Like (Lord of the Rings \green{$\bullet$}) \emph{vs.} Dislike (Harry Potter \& Hobbit \red{$\circ$}) & -0.021 & 0.987 \\
    & Like (Lord of the Rings \green{$\bullet$}) \emph{vs.} Dislike (Lord of the Ring \red{$\bullet$}) & -0.072 & 0.921 \\
    \midrule
    1837 & Like (All \green{$\bullet \circ$}) \emph{vs.} Dislike (All \red{$\bullet$}) & -0.646 & 0.811 \\
    & Like (Training set \green{$\bullet$}) \emph{vs.} Like (Validation set \green{$\circ$}) & 0.627 & -0.532 \\
    & Like (Training set \green{$\bullet$}) \emph{vs.} Dislike (All \red{$\bullet$}) & -0.717 & -0.569 \\
    \bottomrule
  \end{tabular}}
  \label{tab:example_users2}
\end{table*}

In addition, Table~\ref{tab:example_users2} presents the pairwise cosine similarities between the clusters of select movies using the example users and items illustrated in Fig.~\ref{fig:cossim_exapmles}, comparing with the same baseline.
We calculate the cosine similarities between clusters based on the mean of the embeddings belonging to each cluster.

The first example is user 39430, who watched many fantasy movies like the Lord of the Rings, the Hobbit, and the Harry Potter series. This user liked the two Lord of the Rings movies in 2002 and 2003, but disliked an older one released in 1978. According to the third row of Table~\ref{tab:example_users2}, our model successfully captures this difference (low similarity of -0.072), while the baseline model fails to (high similarity of 0.921).
Fig.~\ref{fig:cossim_exapmles} indicates that the non-preferred movies are located far from the preferred ones in the embedding space, with the overall cosine similarity -0.0794.
The baseline model, on the other hand, positions most of these fantasy movies closely to each other, with the cosine similarity 0.980.
Interestingly, our model even distinguishes the disliked movies into two different clusters, clearly characterized by the release years (Fig.~\ref{fig:cossim_exapmles}, left-most).

Another example is user 1837, also used in the Table~\ref{tab:example_users}, who likes Star Wars Episodes IV, V, and VI, but not the Episode I.
Our model embeds the non-preferred series far from the preferred ones with the cosine similarity of -0.646, while CLCRec puts them closer with that of 0.811.
(One might ask why the embedding space in Fig.~\ref{fig:cossim_exapmles} does not reflect this difference. This is because of the dimension reduction to 2D for visualization. The reported cosine similarities are computed with the original embeddings before dimension reduction.)
Comparing the preferred clusters, one from the training and the other from the validation set, we observe that our model locates them closer (0.627) than CLCRec (-0.532).

These examples illustrate that our approach captures the fine-grained difference among the movies of the same genre or even the same series, which is underrepresented with the baseline.
We believe this difference comes from the fact that the existing models train the content feature on classification labels, forced to unlearn subtle differences between items belonging to the same class, while our model is completely free from the classification labels, fully utilizing its capacity to deeply understand the item contents.

\section{Summary}
\label{sec:conclusion}

In this work, we propose a general item content representation learning framework to tackle the item cold-start recommendation problem.
Our proposed framework is agnostic to a specific domain or dataset, applicable to various real-world services with minimal modifications.
Taking advantage of the Transformer architecture, the proposed framework fuses signals from multimodal features in a natural way.
Our framework does not rely on any human-labeled large-scale classification datasets to train modality-specific encoders. Relying solely on user activities, our model learns to represent items preserving fine-grained details of user tastes.
From extensive experiments, we demonstrate the superior performance of our proposed framework both quantitatively and qualitatively, on movie and news domains with multiple datasets.

\section*{Ethical Considerations}

The proposed approach in this paper is about how to better use raw content signals for cold-start item recommendations. We believe the proposed method itself does not impose any immediate positive or negative impact on fairness, privacy, or other ethical concerns, as long as the recommendation systems are trained on fairly collected training data.

We do expect, however, that this line of research possesses a potential to eventually promote user privacy. As the recommendation systems rely more heavily on content signals that are publicly available rather than individual user activity logs, less private data may need to be collected to achieve the same quality of recommendations, Although this work does not claim this line of contributions, it will be an interesting future work to explore and measure how content-based recommendation systems can save the privacy burden from modern recommendation systems.

\bibliographystyle{abbrv}
\bibliography{ref}

\end{document}